\documentclass[twocolumn,superscriptaddress,amsfont,amssymb,amsmath, showpacs,balancelastpage, nofootinbib]{revtex4-1}
\usepackage{graphicx,longtable,mathrsfs,color,array}
\usepackage[unicode=true,pdfusetitle,
bookmarks=true,bookmarksnumbered=true,bookmarksopen=true,bookmarksopenlevel=1,
breaklinks=false,pdfborder={0 0 0},backref=false,colorlinks=true]{hyperref}
\hypersetup{citecolor=blue,filecolor=blue,linkcolor=blue,urlcolor=blue}
\usepackage[usenames,dvipsnames]{xcolor}
\usepackage{amssymb,amsmath,mathtools,mathrsfs,enumitem}
\usepackage{epsfig,subfigure,placeins,float}
\usepackage[update,prepend]{epstopdf} 
\usepackage{booktabs,longtable,multirow} 
\usepackage{exscale,relsize}
\usepackage[normalem]{ulem}
\usepackage[T1]{fontenc}
\usepackage[utf8]{inputenc}
\usepackage{enumerate}
\usepackage{times, mathptmx}
\usepackage{tikz}
\usetikzlibrary{arrows,positioning,decorations.markings,decorations.pathmorphing,calc}

\usetikzlibrary{decorations.pathmorphing}

\usetikzlibrary{arrows,positioning,decorations.markings,decorations.pathmorphing,calc}
\definecolor{hyperref}{RGB}{026,028,087}

\def\gsim{ \lower .75ex \hbox{$\sim$} \llap{\raise .27ex \hbox{$>$}} }
\def\lsim{ \lower .75ex \hbox{$\sim$} \llap{\raise .27ex \hbox{$<$}} }
\def\be{\begin{equation}}
\def\ee{\end{equation}}
\def\bea{\begin{eqnarray}}
\def\eea{\end{eqnarray}}

\newcommand{\Oo}{{{\cal O}(1)}}

\newcommand{\comment}[1]{}

\def\nn{\nonumber}

\def\MP{{M_{\rm P}}}

\definecolor{Gray}{gray}{0.9}
\definecolor{LightCyan}{rgb}{0.88,1,1}

\def\mathcolor#1#{\mathcoloraux{#1}}
\newcommand*{\mathcoloraux}[3]{%
\protect\leavevmode
\begingroup
\color#1{#2}#3%
\endgroup
}
\newlength{\stheight}
\newcommand\textst[1][fu-grey]{
\ifmmode\setlength{\stheight}{+1.0ex}
\else\setlength{\stheight}{+0.5ex}
\fi
\bgroup\markoverwith{\textcolor{#1}{\rule[\the\stheight]{2pt}{1.0pt}}}\ULon
} %
\newcommand{\textins}[2][fu-grey]{
\ifmmode\mathcolor{#1}{#2}
\else\textcolor{#1}{#2}\@\, 
\fi
}
\graphicspath{{./}}
\allowdisplaybreaks

\tikzstyle{vecArrow} = [thick, decoration={markings,mark=at position
1 with {\arrow[semithick]{open triangle 60}}},
double distance=1.4pt, shorten >= 5.5pt,
preaction = {decorate},
postaction = {draw,line width=1.4pt, white,shorten >= 4.5pt}]
\begin{document}

\hypersetup{pageanchor=false} 
\title{Black Hole Ringdown as a Probe for Dark Energy}

\author{Johannes Noller}
\affiliation{Institute for Theoretical Studies, ETH Z\"urich, Clausiusstrasse 47, 8092 Z\"urich, Switzerland}
\affiliation{Institute for Particle Physics and Astrophysics, ETH Z\"urich, 8093 Z\"urich, Switzerland}
\author{Luca Santoni}
\affiliation{Department of Physics, Center for Theoretical Physics, Columbia University, 
538 West 120th Street, New York, New York 10027, USA}
\author{Enrico Trincherini}
\author{Leonardo G. Trombetta}
\affiliation{Scuola Normale Superiore, Piazza dei Cavalieri 7, 56126, Pisa, Italy}
\affiliation{INFN - Sezione di Pisa, 56100 Pisa, Italy}

\begin{abstract}
Under the assumption that a dynamical scalar field is responsible for the current acceleration of the Universe, we explore the possibility of probing its physics in black hole merger processes with gravitational wave interferometers. 
Remaining agnostic about the microscopic  physics, we use an effective field theory approach to describe the scalar dynamics. We investigate the case in which some of the higher-derivative operators, that are highly suppressed on cosmological scales, instead become important on typical distances for black holes. If a coupling to the Gauss-Bonnet operator is one of them, a nontrivial  background profile for the scalar field can be sourced in the surroundings of the black hole, resulting in a potentially large amount of ``hair.'' In turn, this can induce sizeable modifications to the spacetime geometry or a mixing between the scalar and the gravitational perturbations. Both effects will ultimately translate into a modification of the quasinormal mode spectrum in a way that is also sensitive to other operators besides the one sourcing the scalar background.
The presence of deviations from the predictions of general relativity in the observed spectrum can therefore serve as a window onto dark energy physics.
\end{abstract}

\maketitle

\section{Introduction} 

The direct detection of gravitational radiation has marked the birth of gravitational wave astronomy, opening up the possibility to explore the Universe via a new fundamental messenger \cite{LIGOScientific:2018mvr}. This new possibility not only allows us to access the regimes of strong gravity in astronomical systems, but it will also have profound consequences for cosmology. 
Along this line, in the present work we will address the following question: under the assumption that the current acceleration of the Universe is driven by a dynamical scalar sector, to what extent can its physics be probed by looking at the signals emitted in a black hole coalescence process?
Indeed, the presence of a black hole can induce a large pileup effect of the scalar profile in its surroundings, enhancing the field's nonlinearities. This means that some of the Lagrangian operators that provide negligible contributions to the cosmological background may become dominant near the black hole, potentially leaving signatures on the emitted gravitational waves.

For this to be possible, the strong restrictions on the presence of nontrivial scalar profiles around static, spherically symmetric black holes must be overcome. These are usually phrased in terms of so-called `no-hair theorems' (for a review see {\it e.g.} Refs.~\cite{Herdeiro:2015waa, Volkov:2016ehx}). Therefore, in order for the scalar background to be nonzero in the first place and consequently affect the black hole dynamics, the theory must belong to the class of exceptions to such theorems.\footnote{Note that perturbations around the black hole can be affected by the scalar even when its background is vanishing, see {\it e.g.} Refs.~\cite{Tattersall:2017erk, Tattersall:2018nve}. Here we will focus on black holes with nontrivial scalar field backgrounds.} 
The simplest and most studied example to evade the no-hair restrictions has been the case of a linear coupling between the scalar and the Gauss-Bonnet operator (sGB) \cite{Sotiriou:2013qea}. In the literature, such a shift-symmetric operator has so far only been considered in the minimalistic setting in which the only other operator in the Lagrangian for the scalar is the canonical kinetic term \cite{Sotiriou:2014pfa, Ayzenberg:2014aka, Maselli:2015yva, Benkel:2016kcq,Antoniou:2017acq,Blazquez-Salcedo:2018pxo,Witek:2018dmd}. In this paper we will extend this setup substantially. Motivated by our assumption that a shift-symmetric scalar field accounts for the dark energy component of the Universe on cosmological scales, we will consider scalar Lagrangians where, together with the kinetic term and the coupling to Gauss-Bonnet, a very general set of operators is included and study if at least one of them becomes large in the vicinity of a black hole. When this happens, as we discuss below, it has several novel important consequences. {Just to mention two, we will show that both the theoretical \cite{Sotiriou:2014pfa} and observational \cite{Witek:2018dmd} upper bounds on the coupling of the scalar Gauss-Bonnet operator can be relaxed by the presence of an additional term like the cubic Galileon \cite{Nicolis:2008in}, broadening the range of values that such a coupling can take. Moreover, even though the scalar background is sourced by a single operator, at the Schwarzschild radius at least one extra operator becomes of comparable size, opening up a wider spectrum of potentially observable signatures that can serve as a window onto dark energy physics.}

We structure the paper as follows. In Sec.~\ref{sec:setup} we introduce a class of shift-symmetric effective field theories (EFT) that can be responsible for the accelerated expansion of the Universe and we characterize its relevant energy scales. Then we show how the recent measurement of the speed of gravitational waves affects these models. In Sec.~\ref{sec:theory-and-HBH} we discuss the sGB operator that must be present in the EFT to source a background scalar profile around black holes and we estimate the size of the hair. In Sec.~\ref{sec:obs-effects} we derive two classes of observable effects in the ringdown phase due to the dark energy field, namely, deviations from the Schwarzschild geometry and kinetic mixing of metric and scalar perturbations. In Sec.~\ref{sec:constraints} we describe the current most stringent bounds on the size of the hair coming from precision tests of gravity at different length scales. Sec.~\ref{sec:models} contains explicit examples of dark energy models that are consistent with those bounds and, at the same time, give observable effects in a binary black hole coalescence. Finally, Sec.~\ref{sec:conclusions} contains our conclusions.

\section{Setup}\label{sec:setup}

Remaining agnostic about the microscopic theory of the dark sector, we will parametrize the scalar dynamics following an effective field theory perspective. The very general class of (shift-symmetric) dark energy models we will consider in this paper, which was introduced in Ref.~\cite{Pirtskhalava:2015nla}, is schematically defined by a scalar Lagrangian with two energy scales [$L$ is a function with $O(1)$ dimensionless parameters\footnote{Throughout the paper, we are assuming that couplings are $O(1)$, and we do not write them explicitly; we also omit factors of $4\pi$ in all the estimates for simplicity.}],
\begin{equation}\label{pcold}
 \mathcal{L} = \Lambda_2^4  \, L \Big( \frac{(\partial \phi)^2}{\Lambda_2^4} , \frac{\nabla^2 \phi}{\Lambda_3^3} \Big) \, ,
\end{equation}
where $\Lambda_3$ is the UV cutoff of the effective theory and $\Lambda_2 \gg \Lambda_3$, together with $\MP$, is associated with the explicit breaking of the Galileon symmetry $\partial_\mu \phi \to \partial_\mu \phi + b_\mu $ \cite{Nicolis:2008in}. The values of these two energy scales are usually chosen to be such that operators with one derivative per field and the leading higher-derivative (HD) ones, all belonging to the Horndeski class \cite{Horndeski:1974wa}, are similarly responsible for the accelerated expansion of the Universe on cosmological distances. For a time-dependent background of the scalar field $\phi_0(t)$ in a Friedmann-Robertson-Walker (FRW) geometry defined by the Hubble parameter $H(t)$, this assumption implies that $\Lambda_2^4$ is the energy density of the Universe today and therefore $\Lambda_2 = (\MP H_0)^{1/2} \sim 10^7 \, \textup{km}^{-1}$ while $\Lambda_3 = (\MP H_0^2)^{1/3} \sim 10^{-3}\, \textup{km}^{-1}$. In this way, indeed, on the cosmological solution both $X_0 \equiv (\partial \phi_0)^2/\Lambda_2^4$ and $Z_0 \equiv \nabla^2 \phi_0 /\Lambda_3^3 \sim H_0 \partial \phi_0 /\Lambda_3^3$ are of $O(1)$. 

The large hierarchy---10 orders of magnitude---between $\Lambda_2$ and $\Lambda_3$ is necessary to enhance the effect of HD operators up to the point of making them comparable, at the present horizon scale $H_0^{-1}$, with the ones that depend only on the first derivative of the scalar field. Such an extreme regime, which is nevertheless radiatively stable thanks to the approximate Galileon symmetry \cite{Pirtskhalava:2015nla, Santoni:2018rrx}, is mainly motivated by phenomenological reasons: it is the one that allows for the largest variety of potentially observable signatures in the large scale structure. 

The presence of higher-derivative operators at the scale $\Lambda_3$ in the dark energy models described by (\ref{pcold}) is already being strongly constrained by the extraordinarily precise measurement of the speed of gravitational waves, made possible by the observation of the neutron star merger event GW170817 and of its electromagnetic counterpart GRB 170817A \cite{Monitor:2017mdv}.\footnote{Note that the frequencies of the LIGO measurement of GW170817 are close to $\Lambda_3$, so additional assumptions about the UV physics are implicitly made when using this measurement to constrain such ``cosmological'' operators suppressed by $\Lambda_3$ \cite{deRham:2018red}.
}
Consider for example the two types of operators that are present in the EFT (\ref{pcold}) that are usually called quartic and quintic Horndeski.\footnote{The exact definition of the operators is given below in Eq. (\ref{HLag}). Here, we are just keeping track schematically of the number of fields and derivatives.} When evaluated on the FRW background they can affect the speed of propagation of gravity, $c_T$:
\begin{align}\label{ctcontribution}
 \frac{(\partial \phi)^2 (\nabla^2 \phi)^2}{\Lambda_3^6} & \to   \frac{(\partial \phi_0)^4 }{\Lambda_3^6 \MP^2} (\partial h)^2 \nn\\
 \frac{(\partial \phi)^2 (\nabla^2 \phi)^3}{\Lambda_3^9} & \to  \frac{(\partial \phi_0)^4 }{\Lambda_3^6 \MP^2} \frac{\partial^2 \phi_0 }{\Lambda_3^3} (\partial h)^2 \, ;
\end{align}
if such operators play a role in the cosmological evolution, which requires as we discussed $\Lambda_3^3 \MP \sim  \MP^2 H_0^2$, their contribution to $c_T$ is of $O(1)$.   

One way to make the theory consistent with the bound $|c^2_T - 1| \leq 10^{-15}$ is then to assume that the coefficients of all the different operators giving rise to deviations in the speed of propagation of gravitational waves form luminality are extremely small \cite{Baker:2017hug,Creminelli:2017sry,Ezquiaga:2017ekz,Sakstein:2017xjx}.  We will instead assume that the UV cutoff of the dark energy EFT, and therefore the characteristic scale of the higher-derivative operators, is larger than $(\MP H_0^2)^{1/3} \sim 10^{-3}\, \textup{km}^{-1}$ . Let $\Lambda > \Lambda_3$ be this new scale:
\begin{equation}\label{pcnew}
 \mathcal{L} = \Lambda_2^4 \, L \Big( \frac{(\partial \phi)^2}{\Lambda_2^4} , \frac{\nabla^2 \phi}{\Lambda^3} \Big) \, ,
\end{equation}
clearly, the contribution to $c^2_T$ from the two classes of operators in (\ref{ctcontribution}) in this case will be reduced respectively by factors $(\Lambda_3/\Lambda)^6$ and $(\Lambda_3/\Lambda)^9$.  As a result, it is enough that $\Lambda > 10^3 \Lambda_3$ for the theory to be in agreement with observations, by which we mean not only the bound on $|c_T^2-1|$ but also the constraints on graviton decay \cite{Creminelli:2018xsv} and dark energy instabilities induced by gravitational waves \cite{Creminelli:2019kjy}.

There is another independent, and more fundamental, motivation to consider larger values of the scale $\Lambda$. It is well known that general properties of the S-matrix, unitarity, analyticity and crossing symmetry, imply positivity bounds for amplitudes at low energies, which in turn constrain the coefficients of EFT operators \cite{Adams:2006sv}. When these bounds are applied to theories with weakly broken Galileon invariance \cite{Pirtskhalava:2015nla}, they imply that the separation between symmetry-breaking and symmetry-preserving operators cannot be too large while keeping the UV cutoff fixed \cite{Bellazzini:2019xts} (see Refs.~\cite{deRham:2017imi,Bellazzini:2017fep,deRham:2017xox} for closely related prior work). 
More specifically, %
in the case of the Lagrangian (\ref{pcnew}) the condition becomes 
\begin{equation}\label{amplitudes}
\Lambda_{\rm UV} \, \lsim \, (\Lambda/\Lambda_2)^{3/2} \,  10^7 \, {\rm km}^{-1} \, .
\end{equation}
This is to say that, if one separates the scale $\Lambda$ too much from the symmetry-breaking scale $\Lambda_2$, i.e. when the ratio $\Lambda/\Lambda_2$ is taken to be very small, then the new degrees of freedom associated with a UV completion (that respects the basic principles mentioned above) must enter at energies $\Lambda_{\rm UV} < \Lambda$, therefore reducing the regime of validity of the EFT.

As it should be clear from the previous discussion, once the scale $\Lambda$ is taken to be parametrically larger than $(10^{3} \, {\rm km})^{-1}$ there will be no sizable effect on the cosmological evolution from HD operators. From the point of view of dark energy phenomenology, the Lagrangian (\ref{pcnew}) in such a regime is almost indistinguishable from a simple shift-symmetric k-essence model \cite{ArmendarizPicon:2000dh,ArmendarizPicon:2000ah}, which is formally recovered in the limit $\Lambda \to \Lambda_2$. We will argue, however, that the possible existence of higher-derivative operators below $\Lambda_2$ could nevertheless leave an observable imprint. Being irrelevant operators---in the renormalization group (RG)-flow sense---their relative importance grows in the UV, i.e. at shorter distances. Exploiting the new observational window provided by gravitational astronomy, we will discuss in which cases the presence of such interactions can affect the gravitational dynamics at the length scales probed by black hole merger events and in particular during the ringdown phase. 
            
As a consequence, in the following we will be interested in EFTs that are able to describe, together with the evolution of the Universe at cosmological distances, at least black holes of the size probed by LIGO/Virgo, with a characteristic Schwarzschild radius of about $10 \,{\rm km}$. We will therefore consider acceptable theories in which the scale of the UV completion can be as low as
$\Lambda_{UV} \sim 1 \, {\rm km}^{-1}$. On the one hand, according to the condition coming from amplitudes' positivity (\ref{amplitudes}), this requires that the scale suppressing the HD operators satisfies $\Lambda > \Lambda^{pos}_{min} \sim 10^5 \Lambda_3$. On the other hand, purely observational constraints can give a minimum allowed value $\Lambda^{obs}_{min}$ for such a scale, which depending on the model can be either above or below $\Lambda^{pos}_{min}$. The greater of the two should be taken as the most stringent bound, i.e. $\Lambda > max\{\Lambda^{pos}_{min}, \Lambda^{obs}_{min} \}$. See Fig.~\ref{fig-scales} for a schematic representation of the hierarchy of scales for $\Lambda$ discussed above and the various effects taking place at those scales.

\begin{figure*}[t!]
\begin{center}
\begin{tikzpicture}
\draw[->,thick] (0,0) -- (12,0);
\draw[decorate,decoration={snake}] (7,0) -- (11.1,0);
\draw[decorate,decoration={snake}] (4,0) -- (11.1,0);

\foreach \x in {1,4,7,11.1}
\draw[thick] (\x cm,4pt) -- (\x cm,-4pt);

\draw (1,0) node[below=3pt] {$ 10^{-3}\,\textup{km}^{-1} $} node[above=3pt] {$ \Lambda_3 $} node[below=20pt] {HD $\mathcal{O}(1)$} node[below=35pt] {on cosmology};
\draw (4,0) node[below=3pt] {$  $} node[above=3pt] {$ 10^3  \Lambda_3 $} node[below=20pt] {deviation from} node[below=35pt] {$c_T=1$ smaller} node[below=50pt] {than $10^{-15}$};
\draw (7,0) node[below=3pt] {$  $} node[above=3pt] {$ 10^5  \Lambda_3 $} node[below=20pt] {satisfies} node[below=35pt] {amplitudes} node[below=50pt] {bound};
\draw (11,0) node[below=3pt] {$ 10^{7}\,\textup{km}^{-1} $} node[above=3pt] {$ \Lambda_2 = 10^{10}  \Lambda_3 $} node[below=20pt] {conventional} node[below=35pt] {shift-symmetric} node[below=50pt] {EFT};

\draw (4,-0.4) -- (4,-0.6); \draw[->] (4,-0.5) -- (4.5,-0.5);
\draw (7,-0.4) -- (7,-0.6); \draw[->] (7,-0.5) -- (7.5,-0.5);

\end{tikzpicture}
\end{center}

\caption{The hierarchy of scales for $\Lambda$ \eqref{pcnew} and effects of setting $\Lambda$ to the respective scale.
} \label{fig-scales}
\end{figure*}

\section{Shift-symmetric scalar-tensor theories and hairy black holes} \label{sec:theory-and-HBH}

A consistent way to include HD operators in the shift-symmetric dark energy EFT at a scale $\Lambda$ that is much below the scale $\Lambda_2$ suppressing the operators with fewer derivatives, is by doing so in the specific combinations that belong to the shift-symmetric (beyond) Horndeski class \cite{Horndeski:1974wa,Deffayet:2011gz,Gleyzes:2014dya,Zumalacarregui:2013pma}. Indeed, such theories enjoy robust quantum properties due to their weakly broken Galileon invariance \cite{Pirtskhalava:2015nla, Santoni:2018rrx}. Their Lagrangian is
\begin{equation}
S_{H+BH}=\int\mathrm{d}^{4}x\,\sqrt{-g} \sum_{i=2}^{5}{\cal L}_{i}, \label{eq:lagrangian}
\end{equation}
where the ${\cal L}_i$ are functions of the metric $g_{\mu\nu}$ and the derivatives of the scalar field $\phi$. Specifically, we will write them as
\begin{align}\label{HLag}
{\cal L}_{2} & = \Lambda_2^4 \, G_2(X)\,, \nn\\
{\cal L}_{3} & = \frac{\Lambda_2^4}{\Lambda^3} G_{3}(X)\Box\phi\,, \nn \\
{\cal L}_{4}  & = \frac{\Lambda_2^8}{\Lambda^6} G_{4}(X) R - 2 \frac{\Lambda_2^4}{\Lambda^6} G_{4,X}(X)\left[\left(\Box\phi\right)^{2}-\phi_{;\mu\nu}\phi^{;\mu\nu}\right] \nonumber \\
& \quad - \frac{F_4(X)}{\Lambda^6}  {\epsilon^{\mu\nu\rho}}_\sigma\epsilon^{\mu'\nu'\rho'\sigma}\phi_{;\mu}\phi_{;\mu'}\phi_{;\nu\nu'}\phi_{;\rho\rho'}  \, ,\notag \\
{\cal L}_{5} & = \frac{\Lambda_2^{8}}{\Lambda^9} G_{5}(X)G_{\mu\nu}\phi^{;\mu\nu} \nn \\ & \quad +\frac{1}{3} \frac{\Lambda_2^4}{\Lambda^9} G_{5,X}(X)\left[\left(\Box\phi\right)^{3}+2{\phi_{;\mu}}^{\nu}{\phi_{;\nu}}^{\alpha}{\phi_{;\alpha}}^{\mu}-3\phi_{;\mu\nu}\phi^{;\mu\nu}\Box\phi\right]  \nonumber \\
& \quad - \frac{F_5(X)}{\Lambda^9} \epsilon^{\mu\nu\rho\sigma}\epsilon^{\mu'\nu'\rho'\sigma'}\phi_{;\mu}\phi_{;\mu'}\phi_{;\nu\nu'}\phi_{;\rho\rho'} \phi_{;\sigma\sigma'}  \, ,
\end{align}
where $X = g^{\mu\nu} \partial_\mu \phi \partial_\nu \phi/\Lambda_2^4$ is the scalar kinetic term, and the semicolon denotes the covariant derivative. 
Note that not all the functions in ${\cal L}_{4}$ and ${\cal L}_{5}$ are independent, due to the requirement of satisfying the degeneracy conditions to ensure there are only $3$ propagating degrees of freedom.
Here, radiative corrections to the Galileon-breaking operators are suppressed by the ratio $\Lambda^4/\Lambda_2^{4}$.

In order to be able to probe the presence of the HD operators at the distance scales of black hole merger events, there must be substantial deviations of their gravitational dynamics from the prediction of General Relativity in the first place. A condition for this is that a sizable scalar field background, or hair, is sourced by the black holes themselves. This is not a generic feature of shift-symmetric scalar-tensor theories, though. Under some rather strong assumptions, namely staticity, spherical symmetry, asymptotic flatness, and regularity at the horizon, black hole solutions with a nontrivial scalar profile are severely restricted in such theories due to the existence of a no-hair theorem \cite{Hui:2012qt}. The assumption on asymptotic flatness is the first to go in the presence of a cosmological background, but any hair sourced by such background will have negligible effects on astrophysical black holes due to the great separation of the scales involved. Hair can also be generated by the time dependence of the inspiral and merger process. Nevertheless, this kind of hair will not persist in its later stages, i.e., the ringdown, where the black hole can be effectively described as a stationary background plus small perturbations that radiate away. Yet another possible source of hair is rotation, which will be generically present for a black hole produced in a merger process and may in contrast give rise to important effects. However, for slowly rotating black holes, rotation cannot source hair if there was none already in spherical symmetry \cite{Sotiriou:2013qea}, while for rapidly rotating black holes, this remains an open question. In any case, these types of hair do not provide a generic way to probe the presence of HD operators independently of the particular conditions of each event.\footnote{For a review about tests of black hole dynamics in modified theories of gravity, see {\it e.g.} Ref.~\cite{Berti:2018vdi}.}

As a starting point for this kind of analysis, in this paper, we will consider instead a kind of hair which is present even in the very symmetric ideal situation, sourced by the presence of the sGB operator \cite{Sotiriou:2013qea},
\begin{equation}\label{sGB}
 \MP \, \alpha \phi \, \mathcal{R}^2_{GB},
\end{equation}
where $\mathcal{R}^2_{GB}$ is the Gauss-Bonnet invariant
\begin{equation}
 \mathcal{R}^2_{GB} = R_{\mu\nu\rho\sigma} R^{\mu\nu\rho\sigma} -4 R_{\mu\nu} R^{\mu\nu} + R^2.
\end{equation}
We expect our estimates to also hold for slowly rotating black holes \cite{Berti:2013gfa}, while the general situation is beyond the scope of this work. The operator in \eqref{sGB} evades the no-hair theorem by breaking some of its assumptions in a more subtle way.
Note also that it respects the shift symmetry nontrivially, due to the fact that $\mathcal{R}^2_{GB}$ is a total derivative. Given that this operator leads to second-order equations of motion, it must be contained within the Horndeski part of \eqref{eq:lagrangian} ($F_4 = F_5 = 0$). Indeed, it is equivalent to the choice 
\begin{equation}
 G_5 \propto \log(X).
\end{equation}
From the EFT standpoint, the allowed range for the sGB operator coupling $\alpha$ is huge. It is bounded from below by the size of its quantum corrections and from above by the requirement that the strong coupling scale is not below $\Lambda$.\footnote{In Appendix~\ref{app:WBG}  we explain how to derive the upper bound on $\alpha$.} Namely, 
\begin{equation}
 \frac{1}{\MP \, \Lambda} < \alpha < \frac{\MP}{\Lambda^3}.
 \label{boundalpha}
\end{equation}

Hairy solutions in sGB theories have been studied mainly in the case when the only other operators present are the Einstein-Hilbert and the standard kinetic term for the scalar, $X$ \cite{Sotiriou:2014pfa, Ayzenberg:2014aka, Maselli:2015yva, Benkel:2016kcq,Antoniou:2017acq,Blazquez-Salcedo:2018pxo,Witek:2018dmd}. In the language of the above Lagrangian, this case amounts to the choices $G_2 = X$, $G_4 = \Lambda^6/\Lambda_3^6$ and $G_5 = -4  \frac{\Lambda^9}{\Lambda_2^{8}} \MP \, \alpha \log(X)$, with the remaining functions set to zero. Black hole solutions in this context are known to have secondary hair, meaning there is no free parameter, or ``charge,'' associated to them and regular solutions exist only if the coupling $\alpha$ is below a certain threshold \cite{Sotiriou:2014pfa}. It is also relevant to note that in this particular situation, there is no screening mechanism associated with scalar nonlinearities.

In our setup, on the contrary, other operators are present and can actually dominate over the standard kinetic term. Whether this happens or not depends on the size of the background quantities $X_{0}$ and $Z_{0}$, where $Z \equiv \nabla^2 \phi/\Lambda^3$. Starting from asymptotically vanishing values, in the presence of scalar hair, these quantities will grow as one approaches the vicinity of the black hole. However, their maximum values ultimately depend on the size of the sGB coupling $\alpha$. Since we are interested in probing the effect of higher-order operators, we will assume that $\alpha$ is large enough in order to be in a regime where $X_{0} \gg 1$ and $Z_{0} \gg 1$ at the Schwarzschild radius, and possibly farther away. Under this assumption, now suppose that the operator which dominates in this regime (besides the sGB one) has the following power counting:
\begin{equation} \label{large-X}
 \Lambda_2^4 \, G_{m+2}(X)  Z^m \to \Lambda_2^4 \, X^n Z^m.
\end{equation}
Here, $m$ always has to satisfy $m=0,1,2,3$, while $n$ is allowed to be any real number, since we are looking at the large-$X$ asymptotic behavior of the Lagrangian functions $G_i$.\footnote{Once the leading behavior \eqref{large-X} is chosen, this translates into an upper bound on the remaining $G_i$ functions in the large-$X$ limit, such that, on the solution, this assumption remains valid. Furthermore, quantum corrections will also generate for example $\Lambda^4 Z^p$ terms, with $p$ a positive integer. Then, there is also the requirement that these will resum to a function $K(Z)$ which is small enough at large $Z$.} Moreover, we also expect the deviation of the geometry from Schwarzschild, even close to the black hole, not to be very large if one has to be in agreement with current observations \cite{Witek:2018dmd}. %

We now proceed in estimating the size of a background solution for the scalar hair, considering static and spherically symmetric configurations. Under the assumptions stated above, the scalar equation of motion schematically reads\footnote{Note that a contribution like $\partial(\partial\phi X^{n-1}Z^m)$ is also captured by the schematic form given.}
\begin{equation} \label{eom}
 \frac{\Lambda_2^4}{\Lambda^3} \, \partial^2 (X_0^n Z_0^{m-1}) \sim \MP \, \alpha \,  \frac{r_s^2}{r^6},
\end{equation}
where on the right-hand side we are evaluating the Gauss-Bonnet invariant $\mathcal{R}^2_{GB}$ on a Schwarzschild background metric, with $r_s$ the Schwarzschild radius, which acts as the source for the scalar profile at leading order in $\alpha$. Assuming spherical symmetry and a power-law decay for the scalar hair $\phi_0(r)$ [i.e. $\phi_0(r) \sim c_\phi/r^p$, where $c_\phi$ and $p$ are constants], we can easily express the $Z_0$ on the background in terms of $X_0$ as
\begin{equation}
 Z_0 \sim \frac{\Lambda_2^2 X_0^{1/2}}{r \Lambda^3},
\end{equation}
and hence we are able to estimate $X_0$ to be
\begin{equation} \label{bg-X}
 X_0(r)^{n+(m-1)/2} \sim \frac{\MP \, \alpha}{\Lambda_2^2\, r_s r^2} \left( \frac{r \Lambda^3}{\Lambda_2^2} \right)^m,
\end{equation}
where we demand that $\lambda = 2n+m-1 > 0$ and $m \leq 2$, in order for $X_0(r)$ to increase or at most stay constant when moving towards the source. Here we have only kept the slowest decaying part of the solution,  which only takes the present form for black holes, see Appendix \ref{app:non-BH-sources} for other types of sources. Note that, after one includes extra operators besides a standard kinetic term, one can relax the theoretical bound on the size of the sGB operator coming from the regularity of solutions. See Appendix \ref{app:th-bound} for an example with the cubic Galileon.

\section{Observable effects in the ringdown} \label{sec:obs-effects}

One of the main goals of this paper is to show that, even if higher-derivative operators are negligible on cosmological scales, they can nevertheless become larger and possibly be tested at much shorter length scales. A promising opportunity to probe at least some of the self interactions of a scalar field and its coupling to gravity is provided by the observation of gravitational waves emitted during the merger of two black holes \cite{Barack:2018yly}. A robust signal of the presence of an additional degree of freedom can be imprinted on the waves emitted during the ringdown phase, when the newly formed and highly perturbed merger remnant relaxes to its equilibrium configuration. A potential deviation from the predictions of GR can have two origins: the scalar field may have a nontrivial background that deforms the geometry of the final black hole, or there can be a mixing between gravitational and scalar perturbations around the background solution \cite{Berti:2018vdi}. Both will ultimately affect the spectrum of the quasinormal modes (QNM). 
In the following, we will estimate these two effects for a black hole formed in a merger, at a typical distance of the order of the light ring, $r \sim r_s$, where their contribution to the QNM spectrum is the largest. We stress that this will just be a rough estimate of the order of magnitude of these effects. A full computation, though very important and eventually necessary, is beyond the scope of this paper. In fact, these effects have been carefully studied in the particular case of sGB plus canonical kinetic term in {\it e.g.} Refs.~\cite{Blazquez-Salcedo:2016enn, Witek:2018dmd}.

\subsection{Background geometry}
The simplest way to estimate how the presence of a scalar background modifies the spacetime geometry around the black hole, with respect to the Schwarzschild metric, is to compare the sGB operator (\ref{sGB}), evaluated on the unperturbed metric and using the solution (\ref{bg-X}) for the scalar, with $\MP^2$ times the black hole curvature $\mathcal{R} \sim r_s/r^3$ (see Appendix \ref{app:dim-estimates}), as functions of the distance $r$. The ratio between these two quantities will be
\begin{equation}\label{BGdEH}
\varepsilon_0(r) \equiv \frac{\MP\alpha \, \phi_0 \mathcal{R}^2_{GB}}{\MP^2 \mathcal{R}} \sim \alpha \frac{\Lambda_2^2}{\MP r} \sqrt{X_0(r)}.
\end{equation}
Other operators can give contributions with a different $r$-dependence, but which will contribute at the same order of magnitude at $r \sim r_s$. Deviations from the Schwarzschild geometry of ${\cal O} (1)$ are possible in principle in light of the above expression. 
\\

\subsection{Mixing}

The second source of modification for the QNM spectrum is due to the appearance of mixing terms between scalar and gravitational modes in the quadratic Lagrangian expanded around the spherically symmetric background solution. Such terms can be present even if the metric is very close, or exactly equal as in the case of the so-called stealth solutions, to Schwarzschild \cite{Tattersall:2017erk,Tattersall:2019pvx}.

The sGB operator induces a kinetic mixing which schematically has the following form, 
\begin{equation}\label{GBmix}
\MP \alpha \,  \phi \, \mathcal{R}^2_{GB} \supset \alpha \, \frac{r_s}{r^3} \, \partial h_c \partial \pi \equiv \mathcal{Z}^{GB}_{mix} \, \partial h_c \partial \pi,
\end{equation}
where $h_c$ stands for a canonically normalized metric perturbation and $\pi \equiv \phi - \phi_0(r)$. All the other HD operators that appear in (\ref{HLag}) also give rise to a mixing, and in fact one can easily check (see Appendix \ref{app:dim-estimates}) that the contribution from the dominating operator \eqref{large-X} for $m>0$ ---for a regular function $G_i(X)$, $i\geq 3$ ---goes as
\begin{equation} \label{HDmix}
  \mathcal{Z}^{H}_{mix} \sim \frac{r^{2}}{r_s^{2}} {\mathcal Z}^{GB}_{mix},
\end{equation}
on solutions of the equation of motion \eqref{eom}. Note that for $m=0$ only $\mathcal{Z}^{GB}_{mix}$ is present. In what follows we will assume $m>0$, deferring the $m=0$ case to Appendix \ref{app:m0-case}. In any case, at $r \sim r_s$, both contributions to the kinetic mixing are of the same order, but with inequivalent contractions due to their different structure. Presumably, their impact on the QNM spectrum will differ. This will be studied elsewhere. 

To estimate the impact of such a mixing on the ringdown, one has to compare its size to the diagonal elements of the kinetic matrix. As discussed in the previous section, this is where a big difference with respect to most of the literature about Gauss-Bonnet hair appears. If in addition to the sGB operator (\ref{sGB}) only $(\partial \phi)^2$ is present in the scalar Lagrangian, the kinetic term for the perturbation $\pi$ around the background receives no other contributions, and it is therefore canonically normalized. The coefficient ${\mathcal Z}^{GB}_{mix}$ in (\ref{GBmix}) then gives the typical size of the effect. 
In the class of theories considered in this paper, on the other hand, additional operators must be present. Even in the minimal setup, the $G_2(X)$-type operators must be added, because they have to provide the stress-energy tensor responsible for the accelerated expansion of the Universe, together with all the interactions generated by quantum corrections, as required by a consistent EFT description. In this case, the kinetic term for scalar perturbations will be provided with $r$-dependent contributions that grow getting closer to the black hole, $\mathcal{Z}_\pi(r) (\partial \pi)^2$. As we discussed in the previous section when solving the equation of motion, if a value of $r$ is reached such that the dimensionless quantities $Z_0$ and $X_0$ evaluated on the background are much larger than $1$, one can identify the contribution that dominates in this regime and, in this case, estimate the leading correction to the kinetic term, which is given by       
\begin{equation} \label{kinetic-term}
\Delta \mathcal{Z}_\pi \sim X_0^{n-1} Z_0^m \sim \frac{\MP \, \alpha}{\Lambda_2^2} \frac{1}{r_s r^2} \frac{1}{\sqrt{X_0(r)}}.
\end{equation}
When $\mathcal{Z}_\pi \gg 1$, the physical effect of the mixing is obtained only after the scalar perturbation is canonically normalized $\pi = \pi_c/ \sqrt{\mathcal{Z}_\pi}$, and the result reads
\begin{equation} \label{mixing-GB}
 \varepsilon_{mix}(r) \equiv \frac{ \mathcal{Z}^{H}_{mix}}{\sqrt{\mathcal{Z}_\pi}} \sim \sqrt{\alpha} \left( \frac{\Lambda_2^2}{\MP r_s} \right)^{1/2} X_0(r)^{1/4}.
\end{equation}
The same phenomenon, the existence of a large scalar background and, as a consequence, of large corrections to the coefficient of the field perturbation close to massive sources appears in so-called screening mechanisms.\footnote{For a review see {\it e.g.} Refs.~\cite{Khoury:2010xi,Babichev:2013usa}.} In those cases, the field redefinition, which is necessary to canonically normalize the scalar perturbation, produces a suppression of the direct coupling of $\pi$ to matter. The fifth-force exchange of the scalar is thus reduced. For fixed $r_s$, the $r$-dependence of \eqref{mixing-GB} and \eqref{bg-X} indicates that the kinetic mixing effect is maximum at close range to the black hole, i.e. $r \sim r_s$. Moreover, this effect is stronger for smaller black holes, 
\begin{equation}
  \varepsilon_{mix}(r \sim r_s) \propto r_s^{-\frac{(n+1)}{2n+m-1}}.
\end{equation}
A measurement of this effect for various black holes of different masses would allow us to constrain the form of the dominant operator ($n$ and $m$) through the above dependence.

For later use, let us consider two different black holes of Schwarzschild radii $r_{s1}$ and $r_{s2}$, respectively. The ratio of the mixing effects at distances $r_1$ and $r_2$ of each black hole is
\begin{equation} \label{mixing-ratio}
\frac{\varepsilon_{mix}(r_1,r_{s1})}{\varepsilon_{mix}(r_2,r_{s2})} \sim \left[ \left( \frac{r_2}{r_1} \right)^{2-m} \left( \frac{r_{s2}}{r_{s1}} \right)^{2n+m} \right]^{\frac{1}{4n+2(m-1)}},
\end{equation}
where we used Eqs.~\eqref{mixing-GB} and \eqref{bg-X}. Notice that the way this effect scales with distances and masses of the black holes is only dependent on the choice of the dominant operator (i.e. on $n$ and $m$). Other parameters such as the sGB coupling $\alpha$ and the scale $\Lambda$ drop from the above expression.

Another interesting remark is the fact that the sizes of both the effect on the background geometry $\varepsilon_0$ and the effect from kinetic mixing $\varepsilon_{mix}$ are not independent. Indeed, notice that in general at $r \sim r_s$ one has
\begin{equation} \label{relation-effects-rs}
 \varepsilon_0(r_s) = \varepsilon_{mix}(r_s)^2,
\end{equation}
and, therefore, the kinetic mixing effect will always dominate over the effect on the background geometry if both are to be at most of $\mathcal{O}(1)$ at the light ring. %

\section{Constraints from tests of gravity} \label{sec:constraints}

The absolute strength of the effects around black holes discussed above depends on both the choice of coupling $\alpha$ as well as on the form ($n$ and $m$) and the scale $\Lambda$ of the other operators that are present in the Lagrangian. However, the presence of a scalar background may also introduce effects at different scales, where current observations put strong bounds to deviations from GR. Already in LIGO/Virgo events, the absence of an observed dephasing of the gravitational wave signal from the one predicted by GR puts an upper bound on the strength of scalar wave emission \cite{Abbott:2016nmj, Abbott:2017oio}. In a different regime, there are also very precise tests of gravitational physics in the Solar System. One of the strongest bounds of this type comes from Lunar Laser Ranging measurements that put strict constraints on the existence of any kind of fifth force at about the $10^{-10}$ level at distances of the Earth-Moon orbit \cite{Williams:2004qba,Merkowitz:2010kka}. These kinds of bounds will limit the choice of $\alpha$, $\Lambda$ and of the allowed operators. A given choice of $\Lambda$ will furthermore impact the strength of Vainshtein screening and the size of the Vainshtein radii for various systems. It is therefore advisable to revisit situations where this kind of mechanism is necessary in order to agree with observations. We will now discuss these constraints in more detail.

\subsection{Direct scalar-matter coupling}\label{ssec:dc}

When matter is present, it is important to know which is the dominant source for the scalar background. Indeed, besides the sGB operator discussed so far, in general we can expect the scalar to be directly coupled to matter, which can source a scalar profile around matter sources but does not affect the solution around black holes. The sGB operator, instead, sources the scalar in both situations. Let us parametrize the size of such a direct scalar-matter coupling relative to the strength of gravity by $\delta$,
\begin{equation} \label{scalar-matter-coupling}
 \frac{\delta}{\MP} \phi \, T.
\end{equation}
Considering a kinetic mixing of cosmological origin, we expect at least that $\delta > \Lambda_3^3/\Lambda^3$ (Appendix \ref{app:mixing-cosmo}). 
Due to this direct coupling,  a matter source of mass $M_*$ will generate a scalar background with an associated Vainshtein radius of order
\begin{equation} \label{rv-tilde}
 \tilde{r}_v \equiv \frac{1}{\Lambda} \left( \delta \frac{M_*}{\MP} \right)^{1/3} = \frac{\Lambda_3}{\Lambda} \, \delta^{1/3} \, r_v
\end{equation}
where $r_v$ is the usually quoted Vainshtein radius (as sourced by nonlinear interactions suppressed by $\Lambda_3$ in the presence of a $\phi T/\MP$ scalar-matter coupling) \cite{Babichev:2013usa}. The intensity of the screening effect is instead given by the size of $\sqrt{\mathcal{Z}_\pi}$, which grows quickly once inside $\tilde{r}_v$, but it is $\mathcal{O}(1)$ farther away from the source. According to Eq.~\eqref{rv-tilde}, there is in general a much smaller Vainshtein radius compared to the standard case, i.e. $\tilde{r}_v \ll r_v$, and one should check that this does not enter in conflict with current tests of gravity at various scales. Indeed, if screening is needed in order to avoid fifth force constraints, once $\tilde{r}_v$ becomes the size of the system being considered or smaller, one might run into trouble. The way around is to bring $\delta$ down, which although it further decreases $\tilde{r}_v$, also alleviates the problem that screening is trying to solve in the first place.

Consider the smallest value of $\Lambda$ that is generically consistent with $c_T = 1$, i.e. $\Lambda \sim 10^3 \Lambda_3 \sim 1 \, \textup{km}^{-1}$. With this choice and a direct coupling of gravitational strength, $\delta \sim 1$, large systems such as galaxies or galaxy clusters would be in the situation described above, where the fifth force becomes unscreened in their outer regions. This can potentially lead to some tension, and suggests
that $\delta \ll 1$. However, if not of gravitational strength, there is no other well-motivated value for the coupling $\delta$ other than the one generated by kinetic mixing of cosmological origin. With this value of $\Lambda$, we have that $\delta \sim 10^{-9}$ (see Appendix \ref{app:mixing-cosmo}), so we will assume that $\delta$ is approximately of this size.

Now, let us consider the ratio between the source terms of the scalar background for the Earth-Moon system, again assuming that the deviation of the geometry from GR is not larger than $\mathcal{O}(1)$ (and therefore $T \sim \MP^2 \mathcal{R}$),
\begin{equation} \label{GBvsDC}
 \frac{\MP \, \alpha \, \mathcal{R}_{GB}^2}{\delta \,\, \MP \mathcal{R}} \sim \frac{\alpha}{\delta} \frac{r_s^{\oplus}}{r_{E-M}^3}, %
\end{equation}
where $r_s^{\oplus}$ is the Schwarzschild radius of the Earth and $r_{E-M}$ is the typical radius of the orbit of the Moon around the Earth.
If this ratio is equal to or larger than 1,
the background is sourced by sGB also at this scale. In particular, both the estimations for the scalar background and the size of the mixing are then given by the same expressions as for black holes, Eqs.~\eqref{bg-X} and \eqref{mixing-GB}, respectively, appropriately substituting $r_s$ by $r_s^{\oplus}$. We will assume this is the case, and we will later check that this is indeed satisfied when Solar System tests and bounds from amplitudes are taken into account.

\subsection{Scalar wave emission in the inspiral phase}

An important bound comes directly from the effect that a scalar wave emission can have on the inspiral phase of a binary black hole merger. The current best bound on the effective sGB coupling \cite{Witek:2018dmd} at the scales probed during the inspiral, i.e. at $r_{insp}$, comes from the GW151226 event, due to the large number of observed cycles during this phase \cite{Abbott:2016nmj}. In terms of the effective value of the coupling $\alpha_{insp}$ seen by scalar perturbations during this phase, the bound reads
\footnote{Note that the bound from Ref.~\cite{Witek:2018dmd} is obtained using a full simulation of inspiral, merger, and ringdown phases, which is stronger by an order of magnitude than the corresponding pure inspiral constraints \cite{Yagi:2012gp,Yunes:2016jcc}. In an abuse of notation, we will nevertheless label the correspondingly constraint coupling $\alpha_{insp}$ and analogously for related parameters. The presence of HD operators in addition to the sGB interaction will likely affect particularly the highly nonlinear merger phase, potentially altering the value of the bound on $\alpha_{insp}$. Since we are interested in approximate order of magnitude estimates here, we will leave a refinement of our analysis taking into account these effects in more detail for future work and assume $\alpha_{insp}$ can approximately be bounded as discussed above.}
\begin{equation} \label{inspiral-bound}
 \alpha_{insp} \equiv \frac{\alpha}{\sqrt{\mathcal{Z}_\pi(r_{insp})}} < (2.7 \, \textup{km})^2,
\end{equation}
where the denominator accounts for the effect of the Vainshtein screening. Again, this is an important difference with most works that studied observational bounds on the sGB coupling, where usually this effect is not present due to the absence of operators which modify $\mathcal{Z}_\pi$. Nevertheless, here we are only naively estimating how screening will affect the observable coupling, since in dynamical situations such as during a merger it is not yet  clear how effective this mechanism is \cite{deRham:2012fw,Dar:2018dra}. 
Using Eqs.  \eqref{GBmix}, \eqref{mixing-GB}, and \eqref{inspiral-bound}, %
we can then obtain a bound on $\epsilon_{mix}$ (we explicitly evaluate these bounds in Sec.~\ref{sec:models}), where
\begin{equation} \label{emix_insp}
 \varepsilon_{mix}({\rm insp}) = \frac{\alpha_{insp}}{r_s r} \Bigg|_{insp}.
\end{equation}
This can in turn be related to the mixing at the light ring ($r \sim r_s$) of a different black hole with Schwarzschild radius $r_s$ by Eq.~\eqref{mixing-ratio},
\begin{eqnarray} \label{mixing-BH}
 \frac{\varepsilon_{mix}(r_s)}{ \varepsilon_{mix}({\rm insp})} &\sim& \left[ \left( \frac{r_{insp}}{r_s} \right)^{2-m} \left( \frac{r_s^{insp}}{r_s} \right)^{2n+m} \right]^{\frac{1}{4n+2(m-1)}}.
\end{eqnarray}
The bound \eqref{inspiral-bound} then implies a bound on this quantity as well.

\subsection{Solar System tests}

Finally, we now consider constraints coming from highly precise tests of gravity in the Solar System. %
As discussed in the previous section, %
we expect effects from kinetic mixing to dominate observable deviations from GR around black holes. However, in order to avoid violating fifth force constraints, we must check that the same kind of effect is negligible in the Solar System. In particular, at the scale of the Earth-Moon orbit, the mixing must stay below the $10^{-11}$ level, in order to satisfy the Lunar Laser Ranging constraints \cite{Williams:2004qba,Merkowitz:2010kka}. In other words,
\begin{equation} \label{LLR-bound}
 \varepsilon_{mix}(E-M) \, \lsim \, 10^{-11}.
\end{equation}
Here we cannot assume the HD operators dominate over the standard kinetic operator, as it happens in the vicinity of black holes. Indeed, in this scenario screening is irrelevant and, moreover, $\mathcal{Z}^{H}_{mix}$ is subleading compared to $\mathcal{Z}^{GB}_{mix}$ regardless of the value of $m$. Both facts are rooted in the faster decay of the scalar solution around non-black-hole sources, as discussed in Appendix \ref{app:non-BH-sources}. Therefore, the mixing in the Earth-Moon system is simply estimated by
\begin{align} \label{mixing-EM}
\varepsilon_{mix}(E-M) &\sim \alpha \frac{r_{s}^{\oplus}}{r_{E-M}^3}.
\end{align}
While \eqref{inspiral-bound} and \eqref{emix_insp} then bound $\varepsilon_{mix}$ on inspiral scales (as explicitly discussed for GW151226 above), relating this bound to the Solar System constraint \eqref{LLR-bound} via \eqref{mixing-EM} happens through $\alpha$, which however is subject to a certain degree of degeneracy with the parameter $\Lambda$.

\section{Viable models} \label{sec:models}

Now that we have discussed both the observable signatures and constraints, in this section, we proceed to identify explicit models that are viable, i.e. consistent with the above constraints. In what follows, we take the approach to first and foremost maximize the possible observational effects, and then to see which of these models can satisfy the various constraints. Therefore, we will assume that the bound \eqref{inspiral-bound} is saturated. With this, we maximize the size of the kinetic mixing at the light ring of black holes, Eq.~\eqref{mixing-BH}, while its size around the Earth-Moon system, Eq.~\eqref{mixing-EM}, instead, can be used to put an upper bound on $\alpha$ which is model independent.

Saturating the inspiral bound \eqref{inspiral-bound} gives a relation between $\alpha$ and $\Lambda$ for a given choice of operator ($n$ and $m$). Indeed, by using Eqs.~\eqref{mixing-GB} and \eqref{bg-X}, we find that
\begin{equation} \label{alpha-lambda}
\alpha^{2n+m} \Lambda^{3m} \lesssim \frac{\MP^{2n+m-2}}{\Lambda_2^{4(n-1)}} r_{s,insp}^{2n+m} \, r_{insp}^{2-m} \, \varepsilon_{mix}(insp)^{4n+2m-2}. 
\end{equation}
See Appendix \ref{app:m0-case} for the particular case of $m=0$, where $\Lambda$ drops from the equivalent expression and $\alpha$ is fixed directly. 

There is still the freedom to choose the scale $\Lambda$, as long as it is above $10^5 \Lambda_3 \sim 10^2 \textup{km}^{-1}$ in order to satisfy the requirement from amplitudes, Eq.~\eqref{amplitudes}, but still well below $\Lambda_2$.
We also recall that, as discussed in Sec.~\ref{ssec:dc}, the dominant source for the scalar profile in the Solar System is the sGB operator, rather than the direct coupling of cosmological origin, parametrized by $\delta$, since, from \eqref{GBvsDC}, \eqref{LLR-bound} and \eqref{mixing-EM}, the condition
\begin{equation}
 \alpha > \delta \frac{r_{E-M}^3}{r_s^{\oplus}} \sim \left(\frac{\Lambda_3}{\Lambda} \right)^3  \frac{r_{E-M}^3}{r_s^{\oplus}},
\end{equation}
is always satisfied. If for some reason the bound from amplitudes is not to be taken into account, the above condition would give a nontrivial constraint.

\begin{figure*}[t!] 
\begin{center} 
\includegraphics[width=0.4\linewidth]{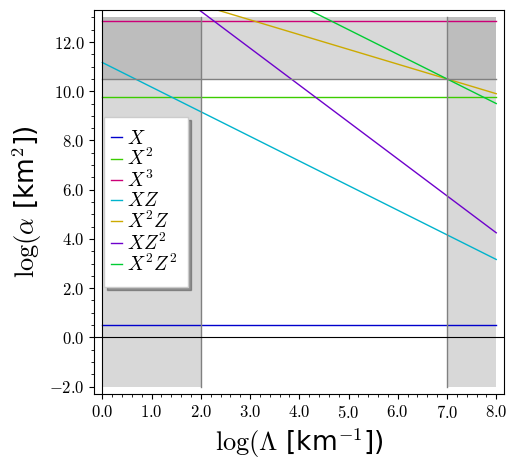}
\end{center}
\caption{
The figure depicts the relationship between $\Lambda$ and $\alpha$, where each line represents the value of $\alpha$ that produces the largest effect during the inspiral phase while remaining within current bounds. The left shaded region is excluded due to positivity bounds. The top shaded region represents where the Lunar Laser Ranging bound \eqref{LLR-bound} is violated. The plot extends to the right until $\Lambda$ of the order $\Lambda_2$. For a given $\Lambda$, values of $\alpha$ below the colored lines are allowed, but they will result in smaller observable effects.
}
\label{fig:plot}
\end{figure*}

Let us now look for explicit examples of models that satisfy all the conditions that were discussed in the previous sections. We take for the radius of the Earth-Moon orbit $r_{E-M} \sim 3\times10^{5} \, \textup{km}$ and the Schwarzschild radius of the Earth $r_s^{\oplus} \sim 10^{-5} \, \textup{km}$, while we use the GW151226 values for the inspiral quantities,\footnote{Actually the inspiral phase spans a range of distances $r_{insp}$ between roughly $3$ and $30$ times the Schwarzschild radius of the black holes, $r_s^{insp}$ \cite{Yunes:2016jcc}. We take an intermediate value which should be acceptable for our purpose of giving order of magnitude estimates for the effects at the light ring.} $r_s^{insp} \sim 30 \, \textup{km}$ and $r_{insp} \sim 300 \, \textup{km}$. This immediately allows us to evaluate a condition
on $\alpha$ which accounts for the Lunar Laser Ranging bound \eqref{LLR-bound}. A complementary condition on these exponents was already mentioned after Eq.~\eqref{bg-X}, namely the requirement that $2n+m-1 > 0$, related to our demand that the background solution $X_0(r)$ decays moving away from the source. We plot the upper bound on $\alpha$ as a function of $\Lambda$ in Fig. \ref{fig:plot} for some example theories defined by a choice of $n$ and $m$, where the shaded regions are excluded.

For these examples, we show in Table \ref{table} the size of the kinetic mixing effect close to both black holes of size $r_s \sim 10 \, \textup{km}$ and $r_s \sim 30 \, \textup{km}$. Recall that, by saturating the bound \eqref{inspiral-bound}, the quoted values for $\varepsilon_{mix}$ are upper bounds for each model. 
\begin{table*}[ht!]
\begin{center}

\begin{tabular}{| c | c | c | c | c | c | c | c |}\hline
$n$ & $m$ & Operator & $\varepsilon_{mix}(r_s \sim 10\text{km})$ & $\varepsilon_{mix}(r_s \sim 30\text{km})$ & $\alpha(\Lambda)$ [$\text{km}^2$] \\\hline
$1$ & $0$ & $X$ & $3 \times 10^{-1}$ & $3 \times 10^{-2}$ & $3$ \\\hline
$2$ & $0$ & $X^2$ & $3 \times 10^{-2}$ & $10^{-2}$ & $10^{10}$ \\\hline
$1$ & $1$ & $XZ$ & $10^{-2}$ & $3 \times 10^{-3}$ & $10^{11} \, (\Lambda \, \text{km})^{-1}$ \\\hline
$2$ & $1$ & $X^2 Z$ & $10^{-2}$ & $3 \times 10^{-3}$ & $3 \times 10^{14} \, (\Lambda \, \text{km})^{-3/5}$ \\\hline
$1$ & $2$ & $X Z^2$ & $10^{-2}$ & $3 \times 10^{-3}$ & $3 \times 10^{16} \, (\Lambda \, \text{km})^{-3/2}$ \\\hline
$2$ & $2$ & $X^2 Z^2$ & $10^{-2}$ & $3 \times 10^{-3}$ & $3 \times 10^{17} \, (\Lambda \, \text{km})^{-1}$ \\\hline
\end{tabular} 
\end{center}
\caption{The table displays a set of possible operators that can dominate the dynamics near a black hole. It includes the maximum value of the kinetic mixing for two scenarios: when the Schwarzschild radius is $r_s =10\,\text{km}$ or $r_s =30\,\text{km}$. Additionally, the corresponding values for $\alpha$ are included, which may depend on the scale $\Lambda$.}
\label{table}
\end{table*}

Notice that for all the models shown in Table \ref{table}, the kinetic mixing effect is always much smaller than 1. According to Eq.~\eqref{relation-effects-rs}, the deviation of the background geometry from Schwarzschild is then even smaller. A question one might ask is whether it is possible to make both these effects to be $\mathcal{O}(1)$ at $r \sim r_s$. While the inspiral bound puts a tight constraint on this possibility, this can nevertheless still be achieved if the kinetic mixing scales steeply enough as one approaches the Schwarzschild radius. From Eq.~\eqref{mixing-ratio}, one can readily see that, fixing all the scales but $r_1=r$, this effect scales as
\begin{equation}
  \varepsilon_{mix}(r) \propto r^{-\frac{(2-m)}{4n+2m-2}}.
\end{equation}
Indeed, one can check that $\mathcal{O}(1)$ effects are possible for $n<1$ for a $r_s = 30 \, \textup{km}$ black hole. Such exotic models can nevertheless be considered acceptable from the EFT point of view, if one intends to remain agnostic about the UV completion of the theory. Indeed, as discussed around Eq.~\eqref{large-X}, in the regime for which $X_0 \gg 1$, there is not necessarily a single operator with an integer value of $n$ that dominates, but rather an infinite tower of operators which collectively show an asymptotic behavior for large $X$ that is compatible with a non integer $n$. 
This means that $\mathcal{O}(1)$ mixing as well as $\mathcal{O}(1)$ deviations from a Schwarzschild background can be achieved with a judicious choice of HD operators in addition to the sGB one, while remaining consistent with all other constraints discussed here.

\section{Conclusions} \label{sec:conclusions}
In this work we have explored the possibility that the dynamics of a scalar field $\phi$ responsible for the accelerated expansion of the Universe can be probed in the strong gravity regime of a black hole coalescence. Our analysis relies on three assumptions: i) the interactions of $\phi$ are shift symmetric; ii) a scalar hair is generated around the final black hole, sourced by a linear coupling between the field and the Gauss-Bonnet operator; and iii) the spin of the black hole is ignored.

Within these conditions, we have found that the presence of certain scalar self-interactions can affect, in an observable way, the spectrum of quasinormal modes emitted during the black hole ringdown. This conclusion is robust, at least from an EFT perspective. The dynamics of the new degree of freedom is parametrized in a general way, based on exact and approximate symmetries which provide well-defined power-counting rules for the derivative and field expansions within the effective Lagrangian. The regime of applicability of the EFT is also imposed to be consistent with the strongest constraints coming from amplitudes' positivity conditions derived up to now in this class of theories \cite{Bellazzini:2019xts}, which also ensures compatibility with constraints on the speed of gravitational waves \cite{Monitor:2017mdv}.    

The details of the resulting deviations from GR predictions are, on the other hand, model dependent, even if they are all ultimately originated by the presence of the sGB coupling. 
This is because at scales of order of the light ring the leading effect, depending on the details of the scalar theory, can be given by different operators. While in the paper we provide only an order of magnitude estimate of such effects, a more complete computation would be useful: the explicit results for QNM spectra obtained so far in the literature are insufficient to fully characterize the potential experimental signatures of this scenario. They are in fact obtained in the limiting case where the only other operator present in the scalar Lagrangian---a part from Gauss-Bonnet---is the kinetic term. 

The variety of possible sources of new effects in the gravitational waveform emitted during the ringdown suggests that, instead of studying each and every case separately, it would be useful to adopt a more model independent approach, like the one recently proposed in \cite{Franciolini:2018uyq}, to compute the QNM spectrum. 

Note that the observable effects discussed here are at the $0.05-0.5$ percent level. While $\Oo$ deviations from Schwarzschild background solutions are already strongly constrained, the sensitivity of current experiments will likely not be enough to probe effects of this size. However, we stress that deviations from GR observable with the next generation of detectors \cite{Punturo:2010zz} are well motivated, since their presence is quite generic and robust.

Finally, we wish to reiterate that the results presented here mean that, in the presence of ``hair,'' the nature of dark energy can be probed with strong gravity observables. While several orders of magnitude separate the scales associated to these regimes, we have shown that a well-defined set of theories is predictive over this range of scales {\it and} yields observable signatures in binary black hole systems.

\subsection*{Acknowledgements}
We acknowledge comments from and useful discussions with P. Creminelli, P. Ferreira, L. Hui, S. Melville, P. Pani, R. Penco, F. Serra, and K. Yagi.
J. N. acknowledges support from Dr. Max R\"ossler, the Walter Haefner Foundation, and the ETH Zurich Foundation. L. S. is supported by Simons Foundation Award No. 555117. E. T. and L. G. T. are supported in part by the MIUR under Contracts No. 2015P5SBHT and No. 2017FMJFMW.

\appendix

\section{Nonrenormalization of Gauss-Bonnet and weakly broken Galileon symmetry} \label{app:WBG} 

The upper bound \eqref{boundalpha} on the coupling $\alpha$ of the sGB operator \eqref{sGB} can be obtained by requiring that the strong coupling scale of the theory is not below $\Lambda$. A simple way to derive it is  by comparing loop diagrams involving the sGB term with tree-level operators in the EFT \eqref{pcnew}. However, in order to get the correct result, one should note that in any quantum loop involving the sGB operator, at the leading order in $1/\MP$ any  scalar leg attached to sGB vertices comes always with at least two derivatives. In other words, the sGB operator satisfies the power counting of the class of operators with  weakly broken Galileon (WBG) symmetry,  introduced in Refs.~\cite{Pirtskhalava:2015nla,Santoni:2018rrx}.\footnote{This is in agreement with the fact that the sGB combination is equivalent to a quintic Horndeski operator, provided a very specific choice of the Horndeski function $G_5 \propto \log (X)$ \cite{Kobayashi:2011nu}.}
 In the following, we provide an explicit check of the latter statement, which in turn will imply that $\alpha$ is bounded from above by $\MP/\Lambda^3$. 

Potentially dangerous contributions, inducing potentially large quantum  corrections to the couplings of the operators of the form $(\partial\phi)^{2n}$, are those coming from sGB  vertices  with two graviton lines, which carry the least suppression in powers of $1/\MP$. We will check that, after tedious integrations by parts, those vertices can in general be rewritten in such a way that the scalar field always carries at least two derivatives,
\begin{equation}
\MP  \alpha \phi \mathcal{R}^2_{GB} \supseteq \frac{\alpha}{\MP} \phi (\partial^2 h_c)^2  \sim \frac{\alpha}{\MP} h_c \partial^2\phi \partial^2 h_c \, ,
\end{equation}
implying therefore that they actually  do not renormalize $(\partial\phi)^{2n}$. To this end, we expand the Riemann tensor, the Ricci tensor and the curvature scalar at linear order in the metric perturbation,
\begin{align}
R_{\mu\nu\rho\sigma} & = \frac{1}{2}\left( \partial_\rho\partial_\nu h_{\mu\sigma} + \partial_\sigma\partial_\mu h_{\nu\rho} - \partial_\sigma\partial_\nu h_{\mu\rho} - \partial_\rho\partial_\mu h_{\nu\sigma} \right) \nn \\ & + \mathcal{O}(h^2) \, ,
\\
R_{\mu\nu} & = \frac{1}{2} \left( \partial_\sigma\partial_\nu h^\sigma_\mu + \partial_\sigma \partial_\mu h^\sigma_\nu - \partial_\mu\partial_\nu h - \partial_\sigma\partial^\sigma h \right) + \mathcal{O}(h^2) \, , 
\\
R & = \partial_\mu\partial_\nu h^{\mu\nu} - \partial_\sigma\partial^\sigma h + \mathcal{O}(h^2) \, .
\end{align}
\begin{widetext}
Plugging this into the definition of the Gauss-Bonnet operator, we obtain
\begin{align}
\phi \mathcal{R}^2_{GB} & =
\phi \left[
\partial_\mu\partial_\nu h_{\rho\sigma} \partial^\mu\partial^\nu h^{\rho\sigma}
+\partial_\mu\partial_\nu h_{\rho\sigma} \partial^\rho\partial^\sigma h^{\mu\nu}
- 2 \partial_\mu\partial_\nu h_{\rho\sigma} \partial^\mu\partial^\rho h^{\nu\sigma}
\right.
 - 2 \partial_\sigma \partial_\nu h^\sigma_{\mu } \partial_\rho \partial^\nu h^{\rho\mu}	
- 2 \partial_\sigma \partial_\nu h^\sigma_{\mu } \partial_\rho \partial^\mu h^{\rho\nu}	 \nn
\\
&\quad + 4 \partial_{\mu}\partial_{\nu} h \partial^\sigma\partial^\mu h^\nu_\sigma + 4 \square h_{\mu\nu} \partial_\sigma\partial^\mu h^{\sigma\nu}
- \partial_\mu\partial_\nu h \partial^\mu\partial^\nu h
- 2 \partial_\mu\partial_\nu h \square h^{\mu\nu} 
-  \square h_{\mu\nu }\square h^{\mu\nu}
\left. + \partial_{\mu}\partial_\nu h^{\mu\nu} \partial_\rho\partial_\sigma h^{\rho\sigma} \right. \nn \\
&\quad \left. - 2 \partial_\mu\partial_\nu h^{\mu\nu} \square h 
+( \square h)^2 \right] \, .
\end{align}
Finally, after straightforward integrating by parts, we find
\begin{align}
\phi \mathcal{R}^2_{GB} & =
\square \phi \partial_\nu h_{\rho\sigma} \partial^\nu h^{\rho\sigma} - \partial_\mu\partial_\nu \phi \partial^\nu h_{\rho\sigma} \partial^\mu h^{\rho\sigma}
- \partial_\rho\partial_\nu \phi \partial_\sigma h^{\sigma\mu} \partial_\mu h^{\nu\rho}  \nn
 +\partial_\rho\partial_\sigma\phi \partial_\nu h^{\sigma\mu}\partial_\mu h^{\rho\nu}
+ \partial_\mu \partial_\nu \phi  \partial_\sigma h^{\sigma\mu } \partial_\rho h^{\rho\nu} 
\\ \nn
&\quad - \partial_\mu\partial_\sigma \phi \partial_\nu h^{\sigma\mu} \partial_\rho h^{\rho\nu} - 2 \partial_\mu \partial^\rho \phi \partial_\mu\partial_\nu h_{\rho\sigma} h^{\nu\sigma}
+ 2 \partial_\nu\partial^\rho\phi \square h_{\rho\sigma} h^{\nu\sigma}
- 2 \partial_\sigma\partial_\nu \phi h^\sigma_\mu \partial^\nu\partial_\rho h^{\rho\mu }
 + 2 \square \phi h^\sigma_\mu \partial_\rho\partial_\sigma h^{\rho\mu } 
\\
&\quad + 2 h \left( \partial_\mu\partial_\nu \phi \partial^\sigma\partial^\mu h^\nu_\sigma - \square \phi \partial_\mu\partial_\nu h^{\mu\nu} \right) - 2 h \partial_\mu\partial_\nu \phi \left(  \square h^{\mu\nu} - \partial_\sigma \partial^\mu h^{\nu\sigma} \right)
- h \left( \partial_\mu\partial_\nu \phi \partial^\mu\partial^\nu h - \square \phi \square h \right) \, ,
\end{align}
which is sufficient to show that any quantum mechanically generated loop correction involving vertices that come from the Gauss-Bonnet operator will not renormalize interactions of the form $(\partial\phi)^{2n}$ at leading order in $\MP$. Corrections instead come with at least an extra suppression in $1/\MP$. 
\end{widetext}

\section{Non-black-hole sources} \label{app:non-BH-sources}

In the case of a shift-symmetric theory, the scalar equation of motion can be written in terms of the divergence of a current $J^{\mu}$
\begin{equation}
 \nabla_\mu J^\mu = M_P \alpha R_{GB}^2.
\end{equation}
Assuming an almost Schwarzschild background, we have
\begin{equation}
 \frac{1}{r^2} \partial_r \left( r^2 J^r \right) = 12 M_P \alpha \frac{r_s^2}{r^6},
\end{equation}
from where, integrating once, we get
\begin{equation}
 J^r = - 3 M_P \alpha \frac{r_s^2}{r^5} + \frac{A}{r^2},
\end{equation}
with $A$ a constant of integration to be fixed by boundary conditions. For a black hole, the requirement of regularity of scalar quantities when $r \to r_s$ demands that $A = 3 M_P \alpha/r_s$, giving rise to the expected fall-off $\phi_0 \sim 1/r$ far away. The situation changes for other objects, where the absence of a horizon instead implies that $A=0$ \cite{Yagi:2015oca}. In such case, the equivalent to Eq. \eqref{bg-X} would be
\begin{equation} 
 X_0(r)^{1/2} + \dots + \left( \frac{\Lambda_2^2}{r \Lambda^3} \right)^m X_0(r)^{n+(m-1)/2} \sim \frac{M_P \alpha }{\Lambda_2^2} \frac{r_s^2}{r^5},
\end{equation}
which implies a much faster fall-off at large distances, $\phi_0 \sim 1/r^4$. 

\section{Black holes with Gauss-Bonnet hair} \label{app:th-bound}

In Ref.~\cite{Sotiriou:2014pfa}, an upper bound on the sGB coupling $\alpha$ has been derived under the assumption that the only operators in the scalar-tensor theory are given by
\begin{equation}\label{t0}
\mathcal{L} = \frac{\MP^2}{2} R - \frac{1}{2}(\partial_\mu\phi)^2 
 + \MP \, \alpha \, \phi \, \mathcal{R}^2_{GB} \, .
\end{equation}
In particular, requiring regularity of the second derivative of the scalar field at the horizon,  the authors of Ref.~\cite{Sotiriou:2014pfa} have shown that $\alpha< \alpha_{max} \equiv r_h^2/\sqrt{192}$, where $r_h$ defines the position of the black hole horizon. 
In the following, we will show that this result is somehow fragile upon deformations of the theory \eqref{t0} and that the bound can indeed be relaxed if other operators become relevant in the vicinity of the black hole.
For simplicity, let us assume that the theory near the horizon is dominated by the following operators,\footnote{In fact, this assumption turns out to be quite general. One can try to consider the more general case of the theory \eqref{HLag} with the function $G_3(X)$ in the form for instance of an arbitrary polynomial of $X$. It turns out that the expression for $\phi''$ near the horizon is dominated by the lowest powers of $X$, leading therefore to the same bound \eqref{eqSot} that we find in the case \eqref{boundatheory}.}
\begin{equation}
\mathcal{L} = \frac{\MP^2}{2} R - \frac{1}{2}(\partial_\mu\phi)^2 
+ \beta (\partial_\mu\phi)^2 \frac{\square\phi}{\Lambda^3}  + \MP \, \alpha \, \phi \, \mathcal{R}^2_{GB} \, ,
\label{boundatheory}
\end{equation}
where we included the cubic Galileon with coupling $\beta$.
Let us parametrize the background metric as follows,
\begin{align}
{\rm d} s^2 = -{\rm e}^{A(r)}{\rm d}t^2 + {\rm e}^{B(r)}{\rm d}r^2 + r^2 \left( {\rm d}\theta^2 +\sin^2\theta \,  {\rm d}\varphi^2  \right) \, ,
\end{align}
and let $r_h$ be the horizon, such that ${\rm e}^{A}\vert_{r\rightarrow r_h^+}\rightarrow0$ and $A'\vert_{r\rightarrow r_h^+}\rightarrow+\infty$.
Solving the $(rr)$-component of the Einstein equations for $B(r)$, in the horizon limit $r\rightarrow r_h^+$, one finds that 
\begin{equation}
{\rm e}^{B}\vert_{r\rightarrow r_h^+}\rightarrow \left( r_h + \frac{4\alpha  \phi'}{\MP}\right)A'\vert_{r\rightarrow r_h^+} \, ,
\label{eB}
\end{equation}
at leading order in $A'\vert_{r\rightarrow r_h^+}$. The presence of the horizon requires that ${\rm e}^{B}$ diverges, %
which translates into the condition $\MP r_h +4\alpha  \phi'(r_h)>0$. Then,  one can plug the result \eqref{eB} into the expression for $\phi''$, which can be obtained for instance from the scalar equation of motion, derived from \eqref{boundatheory}. In the horizon limit, the result takes on the form
\begin{align}
\phi''\vert_{r\rightarrow r_h^+}\rightarrow & -\frac{  12 \alpha   \MP+   r_h^3 \phi '+r_h^2 {\phi'}^2 \left(\frac{4 \alpha}{\MP}  -\frac{\beta }{\Lambda^3}\right)}{\Lambda^3 \MP \left(r_h^4-96 \alpha^2\right)-2 r_h^3 \phi ' \left(\beta  \MP -2 \alpha  \Lambda^3\right)} \nn \\ & \cdot \left(\MP r_h +4 \alpha  \phi'\right) A'\vert_{r\rightarrow r_h^+} \, .
\label{phipp2}
\end{align}
Thus, in order for $\phi''$ to be finite at the horizon, the numerator in \eqref{phipp2} needs to vanish in the limit $r\rightarrow r_h^+$. Solving for $\phi'(r_h)$, one finds two solutions, which are real only if 
\begin{equation}
r_h^4  - 192 \alpha^2 + 48 \beta \frac{\MP \alpha}{\Lambda^3} > 0 \, .
\label{eqSot}
\end{equation}
Setting $\beta=0$, one immediately recovers the upper bound $\alpha_{max} =  r_h^2/\sqrt{192}$ of \cite{Sotiriou:2014pfa}. If instead $\beta\neq0$, Eq.~\eqref{eqSot} allows  a wider range of values for the coupling $\alpha$, provided that $\alpha \beta > 0$. Indeed, assuming $\beta\sim \mathcal{O}(1)$ and $r_h$ in the range of values of standard LIGO/Virgo and LISA black holes, $\alpha$ can now be as large as $\MP /\Lambda^3$, which for $\Lambda \sim 10^2 \text{ km}^{-1}$ corresponds to $\alpha_{max} \sim 10^{32} \text{ km}^2$.

\section{Dimensional estimates for static and spherically symmetric backgrounds} \label{app:dim-estimates}

Here, we present a dimensional estimate of the typical curvature $\mathcal{R}$ outside static spherically symmetric object of mass  $M_*$, assuming the unperturbed GR solution. Since in vacuum both the Ricci tensor and the Ricci scalar are zero, we must look at the full Riemann tensor. A nonvanishing scalar we can build is the Kretschmann scalar,
\begin{equation} \label{RiemannScaling}
 R_{\mu\nu\rho\sigma} R^{\mu\nu\rho\sigma} \simeq \frac{r_s^2}{r^6},
\end{equation}
where we evaluated for the Schwarzschild metric and $r_s = M_*/\MP^2$ is the Schwarzschild radius. From this quantity, we can then give an estimate of the typical curvature as
\begin{equation}
 \mathcal{R} = \sqrt{R_{\mu\nu\rho\sigma} R^{\mu\nu\rho\sigma}} \sim \frac{r_s}{r^3}.
\end{equation}
The Gauss-Bonnet invariant instead, in vacuum, is precisely given by the Kretschmann scalar, and therefore
\begin{equation}
 \mathcal{R}_{GB}^2 \simeq \frac{r_s^2}{r^6}.
\end{equation}

As a worked example, consider the derivation of \eqref{GBmix}. Perturbing the operator $\MP \alpha \,  \phi \, \mathcal{R}^2_{GB}$ around a Schwarzschild background, we obtain the following  quadratic kinetic mixing (modulo numerical factors),
\begin{equation}
\MP \alpha \,  \phi \, \mathcal{R}^2_{GB} \to \MP \alpha R_{\mu\nu\rho\sigma} \partial^\mu \pi \partial^\sigma h^{\nu\rho},
\end{equation}
where the Riemann tensor is taken to be evaluated on the background. Upon normalizing the graviton perturbation ($h \to h_c/M_P$) and explicitly substituting the background scaling for the Riemann tensor \eqref{RiemannScaling}, we then find
\begin{equation}
\MP \alpha R_{\mu\nu\rho\sigma} \partial^\mu \pi \partial^\sigma h^{\nu\rho} \sim \alpha \, \frac{r_s}{r^3} \, \partial h_c \partial \pi,
\end{equation}
reproducing the result of \eqref{GBmix}.
Similarly consider the mixing induced by a HD operator, e.g. the cubic Galileon operator $\Lambda_2^4 X Z$, i.e. $n=1, m=1$ in the notation of \eqref{large-X}. From such a term, we schematically obtain a mixing,
\begin{align}
\frac{1}{\Lambda^3}(\partial \phi_0)^2 \partial \pi \partial h &= \frac{1}{\Lambda_2^4}\left(\frac{\Lambda_3}{\Lambda}\right)^3(\partial \phi_0)^2 \partial \pi \partial h_c \nn \\ &= \left(\frac{\Lambda_3}{\Lambda}\right)^3 X_0 \partial \pi \partial h_c \sim \, \frac{\alpha}{r_s r} \partial \pi \partial h_c,
\end{align}
where we have used \eqref{bg-X} to solve for $X_0$ in the final step, assuming a black hole as the source, and reproduced \eqref{HDmix}.

\section{The $m=0$ case}\label{app:m0-case}

In this appendix we show the alternative expressions of various quantities of interest for the special case $m=0$, always assuming a scalar profile sourced by a black hole. This case is characterized by the absence of higher-derivatives and consequently of any dependence on the scale $\Lambda$. Therefore, as mentioned in the main text, the only source of kinetic mixing will be $\mathcal{Z}^{GB}$ (given by Eq. \eqref{GBmix}). The screening factor instead remains the same as for $m > 0$, namely Eq. \eqref{kinetic-term}. Putting both together we obtain the mixing parameter 
\begin{equation}\label{mixing-GB-m0}
 \varepsilon_{mix}^{(m=0)}(r) = \frac{\mathcal{Z}^{GB}_{mix}}{\sqrt{\mathcal{Z}_\pi}} \sim \sqrt{\alpha} \left( \frac{\Lambda_2^2 \, r_s^3}{\MP r^4} \right)^{1/2} X_0(r)^{1/4}.	
\end{equation}
Notice that the difference between Eqs. \eqref{mixing-GB} and \eqref{mixing-GB-m0} impacts on how we interpret the inspiral bound on $\alpha_{insp}$ in terms of $\varepsilon_{mix}$. Indeed, when $m=0$, in place of Eq. \eqref{emix_insp} we should use
\begin{equation} \label{emix_insp-m0}
 \varepsilon_{mix}^{(m=0)}({\rm insp}) = \alpha_{insp} \, \frac{r_s}{r^3} \Bigg|_{insp}.
\end{equation}

These differences propagate to the way $\alpha$ is bounded, now independently of $\Lambda$, by $\varepsilon_{mix}^{(m=0)}({\rm insp})$. That is, 
\begin{equation} \label{alpha-m0}
\alpha^{2n} \lesssim \frac{\MP^{2n-2}}{\Lambda_2^{4(n-1)}} r_{s,\text{insp}}^{4-6n} \, r_{\text{insp}}^{8n-2} \, \varepsilon_{mix}^{(m=0)}(insp)^{4n-2},
\end{equation}
which it is not just Eq. \eqref{alpha-lambda} evaluated for $m=0$.

\section{Kinetic Mixing from Cosmology}\label{app:mixing-cosmo}

Consider the Lagrangian,
\begin{equation} \label{HD-lagrangian-noGB}
 \mathcal{L} = \MP^2 R + \Lambda_2^4 \, X^n Z^m,
\end{equation}
with $m = 0,1,2,3$ and $n \geq 1$ and $Z = \frac{\partial^2 \phi}{\Lambda^3}$. The HD operators ($m \geq 1$), when expanded around some background solution with $X_0$ and $Z_0$, will generically induce a mixing of the form
\begin{equation}
 \Lambda_2^4 \, X^n Z^m \supset \frac{\Lambda_2^4}{\Lambda^3 \MP} X_0^n Z_0^{m-1} \,  \partial h_c \partial \pi \equiv \mathcal{Z}^{H}_{mix} \, \partial h_c \partial \pi,
\end{equation}
where $\Lambda_2^2 = \MP H_0$. On the other hand, the kinetic term for $\pi$ generically also receives a contribution,
\begin{equation} \label{kinetic-term2}
 \Delta \mathcal{Z}_\pi \sim X_0^{n-1} Z_0^{m}. 
\end{equation}
On the cosmological background, we have $X_0 \sim 1$, and $Z_0 \sim \Lambda_3^3/\Lambda^3$, such that the mixing term and the new contribution to the kinetic term satisfy
\begin{equation} \label{Zmix-cosmo}
 \mathcal{Z}^H_{mix} \sim  \Delta \mathcal{Z}_\pi \sim \left( \frac{\Lambda_3}{\Lambda} \right)^{3m} \ll 1,
\end{equation}
where $\Lambda_3^3 = \MP H_0^2$. Now, after diagonalizing and canonically normalizing, there is an induced coupling with matter of the form
\begin{equation} 
 \frac{1}{\MP} h^c_{\mu\nu} T^{\mu\nu} \to \frac{1}{\MP} \frac{ \mathcal{Z}^H_{mix}}{\sqrt{\mathcal{Z}_\pi}} \pi_c \, T + \dots.
\end{equation}
Assuming there is a standard kinetic term for $\pi$ to begin with, we have $\mathcal{Z}_\pi = 1 + \Delta\mathcal{Z}_\pi \simeq 1$, and then the screening effect can be neglected. Furthermore, if there is at least one operator with $m\geq1$, the induced coupling is at most of order
\begin{equation} \label{matter-scalar-coupling-cosmo}
\left( \frac{\Lambda_3}{\Lambda} \right)^{3} \pi \, T.
\end{equation}

\bibliographystyle{utphys}
\bibliography{bhde}

\end{document}